# Warburg Effect due to Exposure to Different Types of Radiation

*Authors list:* Zhitong Bing[1*], Bin Ao[1*], Yanan Zhang[1*], Fengling Wang[3], Caiyong Ye[1,2], Jinpeng He[1,2], Jintu Sun[1], Jie Xiong[1,2], Nan Ding[1,2], Xiao-fei Gao[1], Ji Qi[1], Sheng Zhang[1], Guangming Zhou[1]† and Lei Yang[1]†

*Authors' Affiliations:* [1]Institute of Modern Physics, Chinese Academy of Sciences, Lanzhou 730000, China

[2]University of Chinese Academy of Sciences, Beijing 100049, China

[3]Biochemistry and molecular laboratory, Medical College of Henan University, Henan 475000, China

* These authors equally contributed to this work.

†*Corresponding author:* Guangming Zhou and Lei Yang

Email: zhougm@impcas.ac.cn & lyang@impcas.ac.cn

Address: 509 Nanchang Rd., Lanzhou 730000, China

**Supplementary Information:** 4 supplement files

1. Supplement Figure: The ATP level, lactic acid level and survival rate by different type of irradiation.

2. Supplementary Dataset 1: The quantitative proteomic data of 2 Gy carbon beam.

3. Supplementary Dataset 2: The quantitative proteomic data of 2 Gy X-ray.

4. Supplementary Dataset 3: The overlap of energy metabolic proteins between 2 Gy carbon beam and 2 Gy X-ray.

**Abstract:** Cancer cells maintain a high level of aerobic glycolysis (the Warburg effect), which is associated with their rapid proliferation. Many studies have reported that the suppression of glycolysis and activation of oxidative phosphorylation can repress the growth of cancer cells through regulation of key regulators. Whether Warburg effect of cancer cells could be switched by some other environmental stimulus? Herein, we report an interesting phenomenon in which cells alternated between glycolysis and mitochondrial respiration depending on the type of radiation they were exposed to. We observed enhanced glycolysis and mitochondrial respiration in HeLa cells exposed to 2-Gy X-ray and 2-Gy carbon ion radiation, respectively. This discovery may provide novel insights for tumor therapy.

# Introduction

Cancer cells demonstrate a high level of aerobic glycolysis even in the presence of ample oxygen, which is called Warburg Effect. Glycolysis, which is highly conserved in evolution process, is the source for ATP and intermediates for amino acids, lipids and nucleotides in proliferating cells [1,2]. Aerobic glycolysis has now been generally accepted as a hallmark of cancer. Therefore, many studies have proposed therapeutics against tumor metabolism by altering aerobic glycolysis to inhibit the growth of cancer cells [3-7]. Many studies have focused on regulators of the Warburg effect, and some regulators have been found to be able to reverse the Warburg effect, which is an important finding for tumor therapy. For example, *PKM2*, *NF-κB*, and *Park2* genes could induce switching from glycolysis to mitochondrial respiration in cancer cells [4,8,9]. Therefore, we assessed whether this switching in cancer cells could be induced by other environmental stimuli.

In this study, we noted an interesting phenomenon wherein cancer cells alternated between glycolysis or mitochondrial respiration depending on the dose and type of radiation. We observed that HeLa cells exposed to 2-Gy X-ray and 2-Gy carbon ion radiation showed enhanced glycolysis and mitochondrial respiration, respectively. A high rate of glycolysis might be associated with cellular radioresistance [10] and hypoxia accelerates the growth of cancer [11]. Even though radiotherapy is one of the major and the most valuable means for cancer treatment, to the best of our knowledge, there is no report on the energy metabolism alternation of cancer cells exposed to ionizing radiation so far.

In this study, a precise quantitative measurement, stable isotope labeling with amino acids in cell culture (SILAC) combined with two-dimensional liquid chromatography-tandem mass spectrometry (2D-LC-MS/MS) shotgun proteomics [12-15], was employed to investigate the response of energy metabolism in HeLa cells exposed to different types of radiation by focusing on metabolism proteins involved in glycolysis, pentose phosphate pathway, tricarboxylic acid (TCA) cycle, oxidative phosphorylation, and fatty acid metabolism. We also measured the levels of ATP and lactic acid with biochemical methods and the expression of key factors with Western blot in HeLa cells after the irradiation exposures. All the results imply that the energy metabolism ways in cancer cells in response to ionizing radiation were radiation type dependent.

## Result

**Protein identification and quantification**

In order to improve the reliability of SILAC experiment, two biological replicates and two technological replicates were applied. With stringent criteria for peptide identification, the false discovery rate (FDR) was controlled below 5%. After filtering using default criteria, 1,088 proteins with two or more assigned unique peptides were identified in 2 Gy carbon-ion irradiated cells (Supplementary Dataset 1) and 1,157 proteins in 2 Gy X-ray irradiated cells (Supplementary Dataset 2). Significant changes in protein expressions were defined as a ratio of ≥ 1.50 or ≤ 0.67. The overlap proteins were classified by PANTHER Classification System ( http://www.pantherdb.org) [16] and the classification results were shown in Figure 1, which indicates that the metabolic process, especially energy metabolism, is intensively influenced by ionizing radiation and many key factors involved in metabolic pathways are up-regulated (Figure 2).

**Data processing**

Metabolic products in cells are useful markers for monitoring cell proliferation in cancer cells [17,18]. Generally, tumor energy source mainly comes from glucose, fatty acid and glutamine [18-21]. According to the KEGG PATHWAY database (http://www.genome.jp/kegg/pathway.html) and the DAVID Bioinformatics Resource 6.7 (http://david.abcc.ncifcrf.gov/home.jsp) [22,23], there are 6 categories for metabolic proteins, glycolysis/gluconeogenesis (GO: 0006096) (GG), pentose phosphate pathway (GO:0006098; GO:0019321) (PPP), tricarboxylic acid cycle (GO:0006099) (TCAC), oxidative phosphorylation (GO:0006119) (OP), glutaminolysis (GO:0006541) (GS) and fatty acid metabolism (GO:0006631) (FAM) [24]. Accordingly, we categorized those proteins identified in the irradiated cells including up-regulated proteins, down regulated proteins and unchanged proteins (Table 1).

As shown in Figure 3 and Supplementary dataset 3, the overlapping proteins among the two kinds of irradiated samples 1 h after exposure to irradiation (2 Gy carbon-ion and 2 Gy X-rays) were involved in all the 6 categories of energy pathways (Figure 3A). We found that the expression of proteins involved in glycolysis were higher in cells irradiated with 2 Gy X-rays than that in cells irradiated with 2 Gy carbon-ion, while the levels of TCA cycle proteins remained unchanged (Figure 3B). Proteins involved in oxidative phosphorylation were increased in both 2 Gy X-ray- and carbon-ion- irradiated samples (Figure 3B). A tendency of increase in the proteins involved in oxidative phosphorylation was observed but there was no significance (Figure 3B and Supplementary dataset 3).

**Expression of energy-metabolism-related proteins**

PKM2 and LDHA play a vital role in aerobic glycolysis within cancer cells. The serine/threonine kinase Akt is closely associated with aerobic glycolysis of cancer cell. Akt stimulates glucose consumption in transformed cell without affecting the rate of oxidative phosphorylation [25]. Besides, recent research reports that the balance between the utilization of glycolysis and mitochondrial respiration is regulated by NF-κB[9]. Therefore, these proteins were selected for validating MS dataset and inferring mechanism of energy metabolic alternation. Western blot analysis (Figure 4) showed that the expression levels of PKM2 and LDHA were basically identical with SILAC results (Table 2). LDHA was significant up-regulated in 2 Gy X-ray-irradiated cells but remained unchanged or down-regulated in 2 Gy X-ray- or carbon-ion-irradiated cells. Akt and NF-κB were transcript factors that were difficult to be detected by MS due to their less abundance. Basing on Western blot analysis we found that Akt was significantly down-regulated in 2 Gy X-ray-irradiated cells but unchanged in 2 Gy carbon-ion-irradiated cells. NF-κB was significantly down-regulated in 2 Gy carbon-ion-irradiated cells but remained unchanged in 2 Gy X-ray-irradiated cells.

**Alternation of ATP level and lactic acid level**

The levels of ATP and lactic acid, the end-products of different energy metabolic pathways, were measured in order to confirm the observations of the changes in energy metabolism in response to ionizing radiation. ATP levels in two different treatments were significantly increased ($p<0.05$) compared to sham control (Figure 5). In the cells irradiated with 2 Gy X-rays, the lactic acid level was significantly increased after irradiation ($p<0.05$), while lactic acid levels in the 2 Gy carbon-ion treatments were not significantly different from the sham control (Figure 6).

# Discussion

The energy metabolism pathways of cancer cell have attracted much attention in cancer therapy research in recent years and several therapeutic strategies have been developed to target energy metabolic pathways by inhibiting glycolysis. In this paper, we revealed that heavy ion beam, an advanced technique for curing cancer, and high dose of X-rays drastically altered the energy metabolism pathways of cancer cells.

Tumor cells show a markedly elevated rate of glycolysis and lactate production, even in the presence of oxygen [26]. Based on our data, most glycolysis proteins significantly increased in cells exposed to 2 Gy X-rays (Figure 3B), however, the expression levels of these proteins remained unchanged in cells exposed to 2 Gy carbon beam (Figure 3B). Moreover, levels of glycolysis proteins in 2 Gy X-ray-irradiated cells showed an increasing trend as compared to those in 2 Gy carbon-ion-irradiated cells. Another report also suggested that increased glycolysis is induced by radiation of 2.5 Gy Gamma ray[27]. Additionally, lactate production in 2 Gy X-ray-irradiated cells was significantly increased as compared to the control (Figure 6). At 1 h post-irradiation, 2 Gy X-rays showed elevated glycolysis and thus increased production of ATP and lactic acid (Figures 3B and 6). But lactic acid levels remained unchanged in the cells exposed to 2 Gy carbon-ion irradiation.

    PKM2 and LDHA are key regulatory enzymes involved in glycolysis in cancer cells. Previous studies show that PKM2 and LDHA expressions are important for tumor growth [4,6]. PFK is associated with pentose phosphate pathway and is essential for tumor growth [28]. In the proteomic and Western blot data of this study, we found that LDHA levels remained unchanged in the cells exposed to 2 Gy carbon beam (Figure 3B and Figure 4), but was up-regulated in the group just exposed to 2 Gy X-ray irradiation (Figure 3B and Figure 4). From Western blot analysis and SILAC experiment, PKM2 has up-regulated trend in two treatments. The pentose phosphate pathway seems to be essential in cancer cells, because of its key role in synthesizing

ribose-5-phosphate (R5P), which is a vital resource for synthesis of nucleotides and nucleic acids [29]. Previous research demonstrates that xylulose-5-phosphate leads to the activation of the transcription factor ChREBP that regulates energy metabolism and fat synthesis [30,31]. Moreover, Transaldolase (TALDO1), glucose-6-phosphate (G6PD) and transketolase (TKT), which are considered to be crucial proteins for cancer cell growth [32], were up-regulated post-irradiation in 2 Gy X-ray-irradiated cells (Figure 3B). Some researcher believed that blocking R5P production could provide a better therapeutic window than that shown by previous anti-metabolic therapies [33]. Additionally, the up-regulated G6P and TKT were necessary for tumor proliferation [34,35]. The result of different change of two treatments in pentose-phosphate pathway indicated that 2 Gy carbon-ion might be better strategy in radiotherapy.

Proteins involved in the TCA cycle and oxidative phosphorylation were mainly up-regulated after irradiation in 2 Gy carbon beam-irradiated cells (Figure 2 and Figure 3B). These pathways are active in the mitochondria. Although it is not clear whether the Warburg effect is induced by mitochondrial dysfunction [6,19,26,36-38], our work showed that at least in HeLa cells the mitochondria can increase ATP production under radiation stress. On the other hand, mitochondrial pyruvate is decarboxlated to acetoin, which inhibits tumor pyruvate dehydrogenase (PDH) [39,40]. It has been described that pyruvate is decarboxylated to an active acetaldehyde through a "non-oxidative" reaction catalyzed by PDH. A decrease in pyruvate oxidation and inhibition of PDH would affect oxidative phosphorylation. We found that the PDHB was remained unchanged at 1 h in all two treatments.

At 1 h after exposure to 2 Gy X-ray irradiation, HeLa cells generated ATP accompanied with increased lactate production. However, HeLa cells exposed to 2 Gy carbon beam did not demonstrate increased lactate production. These data indicated that HeLa cells did not generate more ATP at 1 h after exposure to 2 Gy X-ray irradiation via oxidative phosphorylation. Additionally, the expression of NF-κB was elevated with western blot assay in 2 Gy carbon-ion. In recent report, the researchers discover that NF-κB stimulates oxidative phosphorylation through up-regulation of cytochrome *c* oxidase 2 (SCO2)[9]. We inferred that 2 Gy carbon-ion produced more ATP from oxidative phosphorylation.

Both glucose and glutamine are substrates preferentially consumed by fast-growing tumor cells [41,42]. Previous studies have demonstrated that in HeLa cells, the ATP demand is supported by the aerobic oxidation of both glucose and glutamine [42,43]. In the carbon beam-irradiated cells, the abundant ATP was not derived from aerobic glycolysis, as evidenced by unchanged expression ratios of glycolysis proteins and lactate levels. Although it has been postulated that glutamine is another substrate consumed by fast-growing tumor cells [42,44] and that glutaminolysis is another key metabolic pathway for cell growth and survival [42], many studies have demonstrated that excess glutamine metabolism produces lactate [20,41,45]. Thus, although glutamine would be an important energy substrate for oxidative phosphorylation and related proteins were also up-regulated the end product of glutaminolysis, lactic acid generated by malate dehydrogenase [46,47] was not increased in 2 Gy carbon-ion-irradiated cells. Thus, the increased ATP levels in these treatments were not likely to be derived from glutaminolysis. Previous studies reported that prostate cancer cells enhanced β-oxidation pathways, consuming fatty acids and generating both acetyl-coenzyme A and ATP [19,48]. Fatty acids may be converted into products, which entered the TCA cycle to eventually provide energy for HeLa cells at 1 h after exposure to 2 Gy carbon beam.

Additionally, we investigated the expressions of Akt and NK-κB, which are considered as regulators for the alternation in energy metabolism. It has been reported that the inhibition of Akt could repress tumor growth [35]. We found that Akt was significant down-regulated in 2 Gy X-ray- irradiated cells. Additionally, we observed the significantly up-regulated NK-κB in 2 Gy carbon-ion-irradiated cells (Figures 3A and 3B). NK-κB has been reported as a regulator of mitochondrial respiration [9]. We inferred that the alternation in energy metabolism induced by radiation might be associated with these two regulators. Nevertheless, the correlation between these two factors and the radiation-induced energy metabolic alteration remains unclear and deserves further investigation.

In summary, for the first time we found that HeLa cells exposed to 2 Gy carbon beam did not demonstrate the typical behavior of enhanced glycolysis of generating ATP according to the Warburg effect just like the cells exposed to 2 Gy of X-rays. Both proteomic data and Western blot analysis indicated that the levels of glycolysis-related proteins were mainly unchanged whereas the levels of proteins involved in oxidative phosphorylation were mainly up-regulated. Therefore, we believe that this study reveals a mechanism underlying high outcome of tumor treatment with heavy ion beam and might inspire novel strategy targeting energy generation in tumor radiotherapy. Furthermore, the biological effect of exposure to 2-Gy carbon ion radiation was similar to that of exposure to 4-Gy X-ray radiation (Supplementary Fig). Although the effects of the 2 types of radiation were similar, we found that exposure to 4-Gy X-ray radiation led to more harmful side effects than exposure to 2-Gy carbon ion radiation did.

## Materials and Methods

### Cell culture

Human cervical carcinoma HeLa cell line was kindly provided by Qingxiang Gao (Lanzhou University). The HeLa cells were maintained in RPMI 1640 (Gibco) at 37 ℃ in a 5% $CO_2$ air-humidified incubator. Cells were prepared for irradiation by supplementing the growth medium with light $^{12}C_6^{14}N_4$ L-arginine and $^{12}C_6^{15}N_2$ L-lysine. Control cells were maintained in heavy L-$^{13}C_6^{15}N_4$-arginine and L-$^{13}C_6^{15}N_2$-lysine supplemented medium. At least 7 subcultures were performed to obtain efficiently labeled cell populations. We separated the light labeled cells into 3 groups to be exposed to different levels of irradiation.

**Irradiation**

HeLa cells were trypsinized, counted and seeded in 25 cm$^2$ flasks at a density of 5×10$^5$ cells/flask. After 48 h incubation, sample 1 was irradiated at room temperature with 2 Gy of high-LET carbon beam with original energy of 165 MeV/u generated by the Heavy Ion Research Facility at Lanzhou (HIRFL, Institute of Modern Physics, Chinese Academy of Science). Sample 2 was irradiated with 2 Gy of X-rays at room temperature (Faxitron RX650, Lincolnshire, IL, USA). Cells were returned to the incubator for further incubation.

**Protein extraction and digestion**

Cells were washed twice with PBS and lysed in 8 M urea before sonicated at 4°C. After centrifugation for 30 min at 20,000 g, the supernatants were collected and kept at -80°C for analysis. Protein concentrations were measured using the Bradford method [49].

**2D-LC-MS/MS analysis**

The peptide mixtures were analyzed by 2D-LC coupled to a LTQ-OrbiTrap mass spectrometer (Thermo Electron, San Jose, CA, USA). For each experiment, the peptide mixtures (from 100 μg proteins) were pressure-loaded onto a biphasic silica capillary column (250um id) packed with 3 cm of reverse phase C18 resin (SP-120-3-ODS-A, (3 mm); the Great Eur-Asia Sci & Tech Development, Beijing, China) and 3 cm of Strong cation-exchange resin (Luma 5 um SCX 100A,

Phenomenex, Torrance, CA, USA). The buffers used were 0.1% FA (buffer A), 80% ACN/0.1% FA (buffer B), and 600 mM ammonium acetate/5% ACN/0.1% FA (buffer C).

One MS survey scan, with mass range 400–2000 m/z, was followed by five MS/MS scans. All tandem mass spectra were collected using a normalized collision energy (a setting of 35%), an isolation window of 2 Da, and 1 micro-scan. All searches were performed using a precursor mass tolerance of 3 Da calculated using average isptopic masses, Fixed modification was set for cysteine with the addition of 57.052 Da to represent cysteine carboxyamidation. Variable modification was set for methionine oxidation with the addition of 15.999 Da to represent methionine oxidation. Enzyme cleavage specificity was set to trypsin and no more than two missed cleavages were allowed [49].

**Data analysis and bioinformatics**

Resulting data were collected using the Xcalibur data system (ThermoElectron, Waltham, MA, USA) and was interpreted using the SEQUEST algorithm of Bioworks 3.3.1 (Thermo Fisher Scientific, USA) against the NCBI RefSeq database (2008, 5, 18). Filtering of peptide identifications in Bioworks were set as follows; Xcorr≥1.8 (z = 1), 2.2 (z = 2), 3.5 (z = 3), Sp≥ 500, Rsp≥ 5, proteins with number of peptide≥2 and consensus score≥10. Subsequent analysis included assigning peptides to the spectra, validating the peptide assignments to remove incorrect results, determining relative quantitation ratios between heavy and light isotope labeling, and inferring protein identifications from the assigned peptides. The open source tools called TPP (Version 4.4.1) was used to validate peptides assigned to MS/MS spectra (PeptideProphet) [50] and to infer the identity of proteins in samples with differentially labeled proteins (ProteinProphet) [51] Minimum peptide length was considered to be 7 and results below a PeptideProphet probability of 0.95 were not considered further.

For protein quantitation, only unmodified peptides and peptides modified by acetyl (protein N terminus) and oxidation (Met) were used. Thus, peptides are not used for quantitation at the protein level. If a counterpart to phosphorylated peptide was identified, this counter peptide was also not used for protein quantitation.

The annotations and functions of proteins were obtained from DAVID Bioinformatics Resources 6.7 (http://david.abcc.ncifcrf.gov/home.jsp)[22,23]. For unknown proteins, annotations were performed by searching the Uniprot and NCBI protein databases. We classified these proteins according to the Biological Process Ontology guidelines of the Gene Ontology project. MS data were searched using SEQUEST algorithm (Ver. 2.8) against the human database, which was released on May 27, 2008, and contains 37,869 protein sequences. The database was reversed and attached to estimate the false discovery rate (FDR).

**ATP measurements**

Following irradiation and subsequent incubation for 1 h at 37°C, cells were washed thoroughly with 0.9 % sodium chloride solution, harvested by centrifugation, resuspended in distilled water and then lysed in ice water using an ultrasonic cell disrupter (Sonics, Newtown, CT, USA). Sonication was performed 4 times for 10 s each time with a 30 s pause between sonication bursts. Then, the lysate was boiled for 10 min in a boiling water bath, cell debris was removed by centrifugation at 4000 rpm for 10 min, and the ATP levels in the supernatant were measured using an ATP determination kit (Nanjing Jiancheng, Nanjing, China). The total protein concentration in the cell lysates was assayed using a BCA protein assay kit (Pierce, Rockford, IL, USA).

**Lactic acid measurements**

Lactate production was measured using an enzymatic kit (Nanjing Jiancheng) by following the manufacturer's instruction. These results were normalized by cell counts. Briefly, $NAD^+$ was added to media and stoichiometrically converted to NADH by lactate in the media. The levels of NADH were then quantified colorimetrically, as described by the manufacturer.

**Western Blot analysis**

A standard protocol as described [52] was followed. Antibodies to pyruvate kinase isozyme M2 (PKM2) (sc-365684), nuclear factor kappa B (NK-κB) (sc-33039) and lactate dehydrogenase A (LDHA) (sc-27230) were purchased from Santa Cruz Biotechnology. Anti-AKT (9272) antibodies were purchased from Cell Signaling Technology.

**Acknowledgments**

We thank the HIRFL-CSR crew for providing the carbon ion beam. This work was supported by grants from the Major State Basic Research Development Program of China (973 Program, No. 2010CB834201), the Hundred Talent Program of the Chinese Academy of Sciences (No. 0760140BR0), the "Strategic Priority Research Program" of the Chinese Academy of Sciences (No. XDA01040411 & XDA01020304), the National Natural Science Foundations of China awarded to Guangming Zhou (No. 10979062) and the Foundation for Young Talents of Gansu (No.1208RJYA013)

# Reference

1       Fothergillgilmore, L. A. The evolution of the glycolytic pathway. *Trends in Biochemical Sciences* **11**, 47-& (1986).

2       Murphy, J. P. & Pinto, D. M. Targeted Proteomic Analysis of Glycolysis in Cancer Cells. *J. Proteome Res.* **10**, 604-613, doi:10.1021/pr100774f (2011).

3       Pan, J. G. & Mak, T. W. Metabolic Targeting as an Anticancer Strategy: Dawn of a New Era? *Sci. STKE* **2007**, pe14-, doi:10.1126/stke.3812007pe14 (2007).

4       Christofk, H. R. *et al.* The M2 splice isoform of pyruvate kinase is important for cancer metabolism and tumour growth. *Nature* **452**, 230-U274, doi:10.1038/nature06734 (2008).

5       Pitroda, S. P. *et al.* STAT1-dependent expression of energy metabolic pathways links tumour growth and radioresistance to the Warburg effect. *Bmc Medicine* **7**, doi:10.1186/1741-7015-7-68 (2009).

6       Fantin, V. R., St-Pierre, J. & Leder, P. Attenuation of LDH-A expression uncovers a link between glycolysis, mitochondrial physiology, and tumor maintenance. *Cancer Cell* **9**, 425-434, doi:DOI: 10.1016/j.ccr.2006.04.023 (2006).

7       Pelicano, H., Martin, D. S., Xu, R. H. & Huang, P. Glycolysis inhibition for anticancer treatment. *Oncogene* **25**, 4633-4646 (2006).

8       Zhang, C. *et al.* Parkin, a p53 target gene, mediates the role of p53 in glucose metabolism and the Warburg effect. *Proceedings of the National Academy of Sciences* **108**, 16259-16264, doi:10.1073/pnas.1113884108 (2011).

9       Mauro, C. *et al.* NF-[kappa]B controls energy homeostasis and metabolic adaptation by upregulating mitochondrial respiration. *Nat Cell Biol* **13**, 1272-1279, doi:http://www.nature.com/ncb/journal/v13/n10/abs/ncb2324.html#supplementary-information (2011).

10      Sattler, U. G. A. *et al.* Glycolytic metabolism and tumour response to fractionated irradiation. *Radiotherapy and Oncology* **94**, 102-109, doi:10.1016/j.radonc.2009.11.007 (2010).

11      Cui, J., Mao, X., Olman, V., Hastings, P. J. & Xu, Y. Hypoxia and miscoupling between reduced energy efficiency and signaling to cell proliferation drive cancer to grow increasingly faster. *Journal of Molecular Cell Biology* **4**, 174-176, doi:10.1093/jmcb/mjs017 (2012).

12      Ong, S. E. *et al.* Stable isotope labeling by amino acids in cell culture, SILAC, as a simple and accurate approach to expression proteomics. *Molecular & Cellular Proteomics* **1**, 376-386,

doi:10.1074/mcp.M200025-MCP200 (2002).

13    Ong, S. E. & Mann, M. Mass spectrometry-based proteomics turns quantitative. *Nature Chemical Biology* **1**, 252-262 (2005).

14    Mann, M. Functional and quantitative proteomics using SILAC. *Nature Reviews Molecular Cell Biology* **7**, 952-958, doi:10.1038/nrm2067 (2006).

15    Chen, E. I. & Yates, J. R. Cancer proteomics by quantitative shotgun proteomics. *Molecular Oncology* **1**, 144-159, doi:10.1016/j.molonc.2007.05.001 (2007).

16    Thomas, P. D. *et al.* PANTHER: a browsable database of gene products organized by biological function, using curated protein family and subfamily classification. *Nucleic Acids Research* **31**, 334-341, doi:10.1093/nar/gkg115 (2003).

17    Hsu, P. P. & Sabatini, D. M. Cancer cell metabolism: Warburg and beyond. *Cell* **134**, 703-707, doi:10.1016/j.cell.2008.08.021 (2008).

18    Heiden, M. G. V., Cantley, L. C. & Thompson, C. B. Understanding the Warburg Effect: The Metabolic Requirements of Cell Proliferation. *Science* **324**, 1029-1033, doi:10.1126/science.1160809 (2009).

19    Moreno-Sanchez, R., Rodriguez-Enriquez, S., Marin-Hernandez, A. & Saavedra, E. Energy metabolism in tumor cells. *Febs Journal* **274**, 1393-1418, doi:10.1111/j.1742-4658.2007.05686.x (2007).

20    Friday, E., Oliver, R., Welbourne, T. & Turturro, F. Glutaminolysis and glycolysis regulation by troglitazone in breast cancer cells: Relationship to mitochondrial membrane potential. *Journal of Cellular Physiology* **226**, 511-519, doi:10.1002/jcp.22360 (2011).

21    Kuhajda, F. P. Fatty Acid Synthase and Cancer: New Application of an Old Pathway. *Cancer Research* **66**, 5977-5980, doi:10.1158/0008-5472.can-05-4673 (2006).

22    Huang, D. W., Sherman, B. T. & Lempicki, R. A. Systematic and integrative analysis of large gene lists using DAVID bioinformatics resources. *Nat. Protocols* **4**, 44-57, doi:http://www.nature.com/nprot/journal/v4/n1/suppinfo/nprot.2008.211_S1.html (2008).

23    Huang, D. W., Sherman, B. T. & Lempicki, R. A. Bioinformatics enrichment tools: paths toward the comprehensive functional analysis of large gene lists. *Nucleic Acids Research* **37**, 1-13, doi:10.1093/nar/gkn923 (2009).

24    Resendis-Antonio, O., Checa, A. & Encarnacion, S. Modeling Core Metabolism in Cancer Cells: Surveying the Topology Underlying the Warburg Effect. *Plos One* **5**, Article No.: e12383, doi::10.1371/journal.pone.0012383 (2010).

25    Elstrom, R. L. *et al.* Akt Stimulates Aerobic Glycolysis in Cancer Cells. *Cancer Research* **64**, 3892-3899, doi:10.1158/0008-5472.can-03-2904 (2004).

26    Warburg, O. On the origin of cancer cells. *Science* **123**, 309-314 (1956).

27  Sriharshan, A. *et al.* Proteomic analysis by SILAC and 2D-DIGE reveals radiation-induced endothelial response: Four key pathways. *Journal of Proteomics* **75**, 2319-2330, doi:10.1016/j.jprot.2012.02.009 (2012).

28  Chesney, J. *et al.* An inducible gene product for 6-phosphofructo-2-kinase with an AU-rich instability element: Role in tumor cell glycolysis and the Warburg effect. *Proceedings of the National Academy of Sciences* **96**, 3047-3052, doi:10.1073/pnas.96.6.3047 (1999).

29  Ramos-Montoya, A. *et al.* Pentose phosphate cycle oxidative and nonoxidative balance: A new vulnerable target for overcoming drug resistance in cancer. *International Journal of Cancer* **119**, 2733-2741, doi:10.1002/ijc.22227 (2006).

30  Kabashima, T., Kawaguchi, T., Wadzinski, B. E. & Uyeda, K. Xylulose 5-phosphate mediates glucose-induced lipogenesis by xylulose 5-phosphate-activated protein phosphatase in rat liver. *P Natl Acad Sci USA* **100**, 5107-5112, doi:10.1073/pnas.0730817100 (2003).

31  Iizuka, K. & Horikawa, Y. ChREBP: A glucose-activated transcription factor involved in the development of metabolic syndrome. *Endocrine Journal* **55**, 617-624 (2008).

32  Cascante, M. *et al.* Metabolic control analysis in drug discovery and disease. *Nature Biotechnology* **20**, 243-249 (2002).

33  Tennant, D. A., Duran, R. V. & Gottlieb, E. Targeting metabolic transformation for cancer therapy. *Nat Rev Cancer* **10**, 267-277 (2010).

34  Coy, J. F., Dressler, D., Wilde, J. & Schubert, P. Mutations in the transketolase-like gene TKTL1: Clinical implications for neurodegenerative diseases, diabetes and cancer. *Clinical Laboratory* **51**, 257-273 (2005).

35  Tennant, D. A., Duran, R. V. & Gottlieb, E. Targeting metabolic transformation for cancer therapy. *Nature Reviews Cancer* **10**, 267-277, doi:10.1038/nrc2817 (2010).

36  Verma, M., Kagan, J., Sidransky, D. & Srivastava, S. Proteomic analysis of cancer-cell mitochondria. *Nat Rev Cancer* **3**, 789-795 (2003).

37  Altenberg, B. & Greulich, K. O. Genes of glycolysis are ubiquitously overexpressed in 24 cancer classes. *Genomics* **84**, 1014-1020, doi:10.1016/j.ygeno.2004.08.010 (2004).

38  Weinhouse, S. Warburg hypothesis 50 years later. *Zeitschrift Fur Krebsforschung Und Klinische Onkologie* **87**, 115-126 (1976).

39  Baggetto, L. G. & Lehninger, A. L. Formation and utilization of acetoin, an unusual product of pyruvate metabolism by ehrlich and AS30-D tumor mitochondria. *Journal of Biological Chemistry* **262**, 9535-9541 (1987).

40  Baggetto, L. G. Deviant energetic metabolism of glycolytic cancer-cells. *Biochimie* **74**, 959-974, doi:10.1016/0300-9084(92)90016-8 (1992).

41  Piva, T. J. & McEvoy-Bowe, E. Oxidation of glutamine in HeLa cells: Role and control of truncated TCA cycles in tumour mitochondria. *Journal of*

*Cellular Biochemistry* **68**, 213-225 (1998).

42	Reitzer, L. J., Wice, B. M. & Kennell, D. Evidence that glutamine, not sugar, is the major energy source for cultured HeLa cells. *Journal of Biological Chemistry* **254**, 2669-2676 (1979).

43	Rodriguez-Enriquez, S. *et al.* Control of cellular proliferation by modulation of oxidative phosphorylation in human and rodent fast-growing tumor cells. *Toxicology and Applied Pharmacology* **215**, 208-217, doi:10.1016/j.taap.2006.02.005 (2006).

44	Board, M., Humm, S. & Newsholme, E. A. Maximum activities of key enzymes of glycolysis, glutaminolysis, pentose-phosphate pathway and tricarboxylic-acid cycle in normal, neoplastic and suppressed cells. *Biochemical Journal* **265**, 503-509 (1990).

45	Wise, D. R. *et al.* Myc regulates a transcriptional program that stimulates mitochondrial glutaminolysis and leads to glutamine addiction. *Proceedings of the National Academy of Sciences* **105**, 18782-18787, doi:10.1073/pnas.0810199105 (2008).

46	Dang, C. V. PKM2 Tyrosine Phosphorylation and Glutamine Metabolism Signal a Different View of the Warburg Effect. *Sci. Signal.* **2**, pe75-, doi:10.1126/scisignal.297pe75 (2009).

47	Moreadith, R. W. & Lehninger, A. L. The pathways of glutamate and glutamine oxidation by tumor cell mitochondria. Role of mitochondrial NAD(P)+-dependent malic enzyme. *Journal of Biological Chemistry* **259**, 6215-6221 (1984).

48	Samudio, I., Fiegl, M. & Andreeff, M. Mitochondrial Uncoupling and the Warburg Effect: Molecular Basis for the Reprogramming of Cancer Cell Metabolism. *Cancer Research* **69**, 2163-2166, doi:10.1158/0008-5472.can-08-3722 (2009).

49	Li, J. *et al.* Proteomic analysis of mitochondria from Caenorhabditis elegans. *Proteomics* **9**, 4539-4553, doi:10.1002/pmic.200900101 (2009).

50	Han, D. K., Eng, J., Zhou, H. L. & Aebersold, R. Quantitative profiling of differentiation-induced microsomal proteins using isotope-coded affinity tags and mass spectrometry. *Nature Biotechnology* **19**, 946-951 (2001).

51	Nesvizhskii, A. I., Keller, A., Kolker, E. & Aebersold, R. A statistical model for identifying proteins by tandem mass spectrometry. *Analytical Chemistry* **75**, 4646-4658, doi:10.1021/ac0341261 (2003).

52	He, J. P. *et al.* Cell cycle suspension A novel process lurking in G(2) arrest. *Cell Cycle* **10**, 1468-1476, doi:10.4161/cc.10.9.15510 (2011).

## Authors' contributions

ZTB, BO, GMZ and LY made substantial contribution to conception, design, acquisition, interpretation of the data, as well as drafting and approving the final manuscript. YNZ prepared figure 4(A,B). FLW, CYY and JTS analyzed the data of MS. JPH and JX made substantial contributions to material preparation and final approval of the manuscript. ND and SZ participated in modification and discussion in manuscript. All authors reviewed the manuscript.

## Additional Information

Competing financial interests
The authors declare no competing financial interests

# Figure Legends

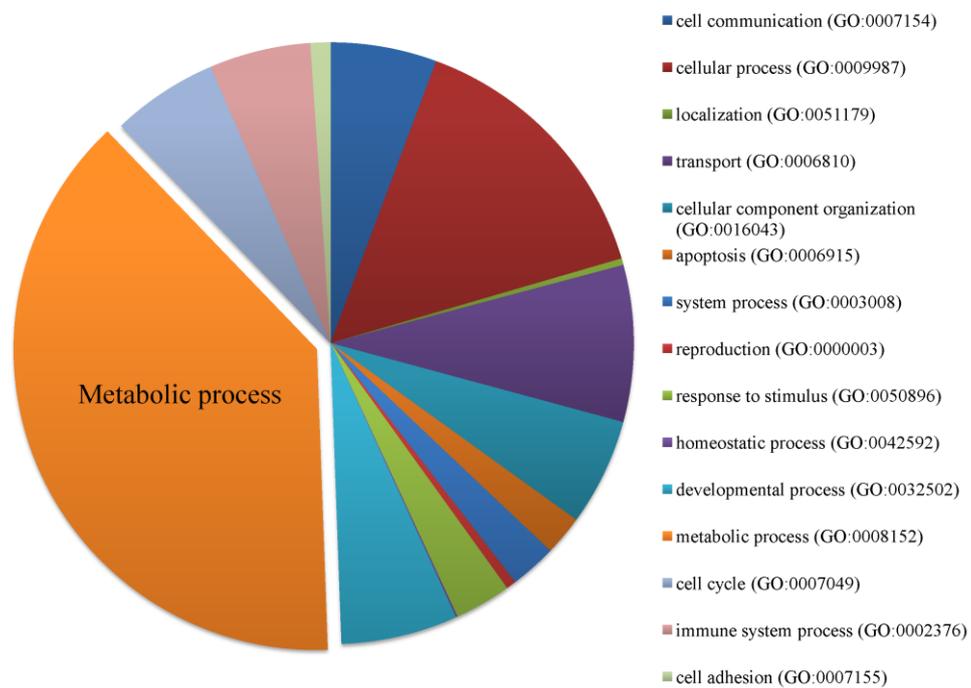

**Figure 1. Biological distribution of the overlapping proteins.**

The pie chart shows the distribution of biological processes to the overlap of three samples using PANTHER classification system. The metabolic process plays a dominant role in three samples.

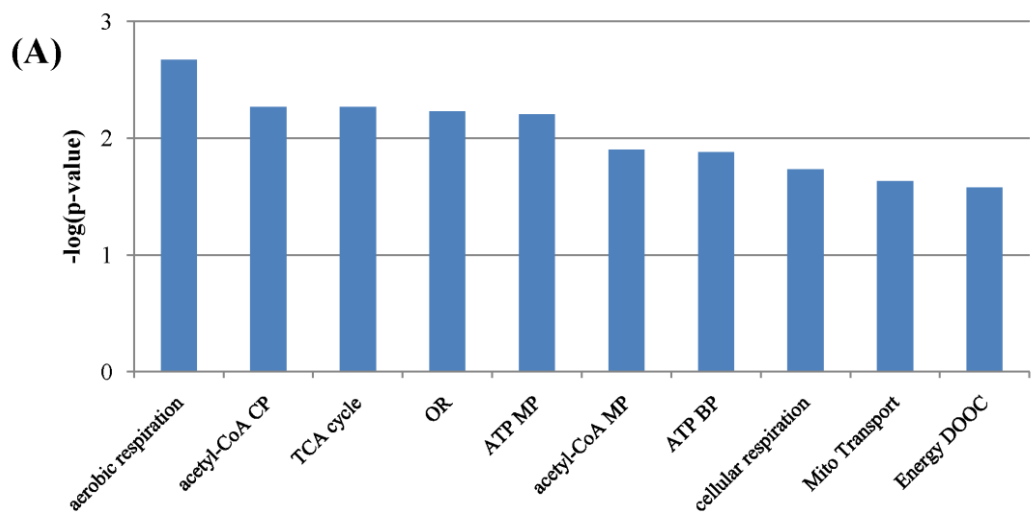

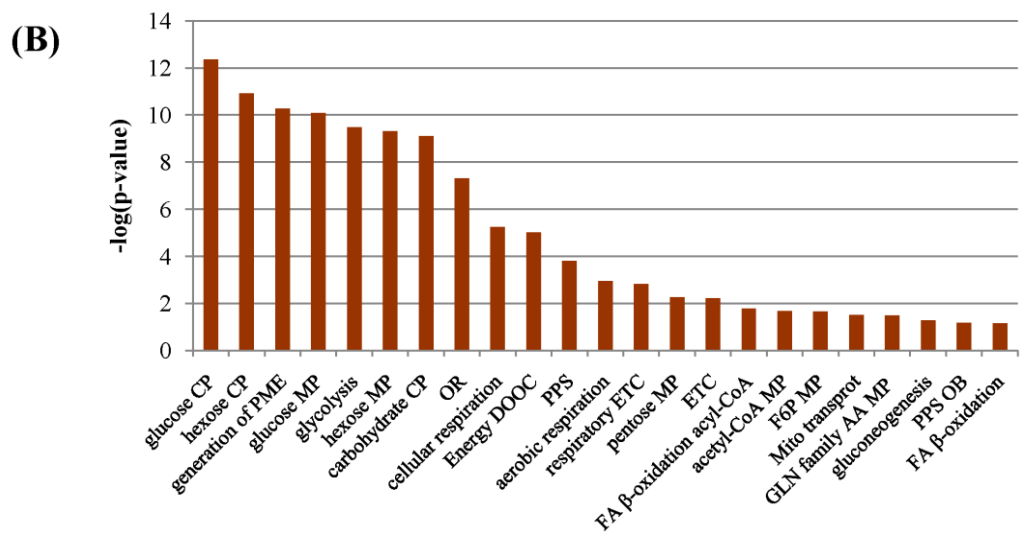

**Figure 2**. **The analysis of up-regulated proteins in energy metabolism at 1h after irradiation by 2 Gy carbon beam**

(A); 2 Gy X-rays (B);. The classification used DAVID Bioinformatics Resources 6.7. (-log(p-value)>1.5 is significant). The biological process abbreviations were following: acetyl-CoA catabolic process (acetyl-CoA CP), tricarboxylic acid cycle (TCA cycle), oxidation reduction (OR), ATP metabolic process (ATP MP), ATP biosynthetic process (ATP BP), mitochondrial transport (Mito Transport), energy derivation by oxidation of organic compounds (Energy DOOC), glucose catabolic process (glucose CP), hexose catabolic process (hexose CP), generation of precursor metabolites and energy (generation of PME), glucose metabolic process (glucose MP), hexose metabolic process (hexose MP), carbohydrate catabolic process (carbohydrate CP), pentose-phosphate shunt (PPS), respiratory electron transport chain (respiratory ETC), pentose metabolic process (pentose MP), electron transport chain (ETC), fatty acid beta-oxidation using acyl-CoA oxidase (FA β-oxidation acyl-CoA), acetyl-CoA metabolic process (acetyl-CoA MP), fructose 6-phosphate metabolic process (F6P MP), glutamine family amino acid metabolic process (GLN family AA MP), pentose-phosphate shunt, oxidative branch (PPS OB), fatty acid beta-oxidation (FA β-oxidation), carboxylic acid catabolic process (CA CP)

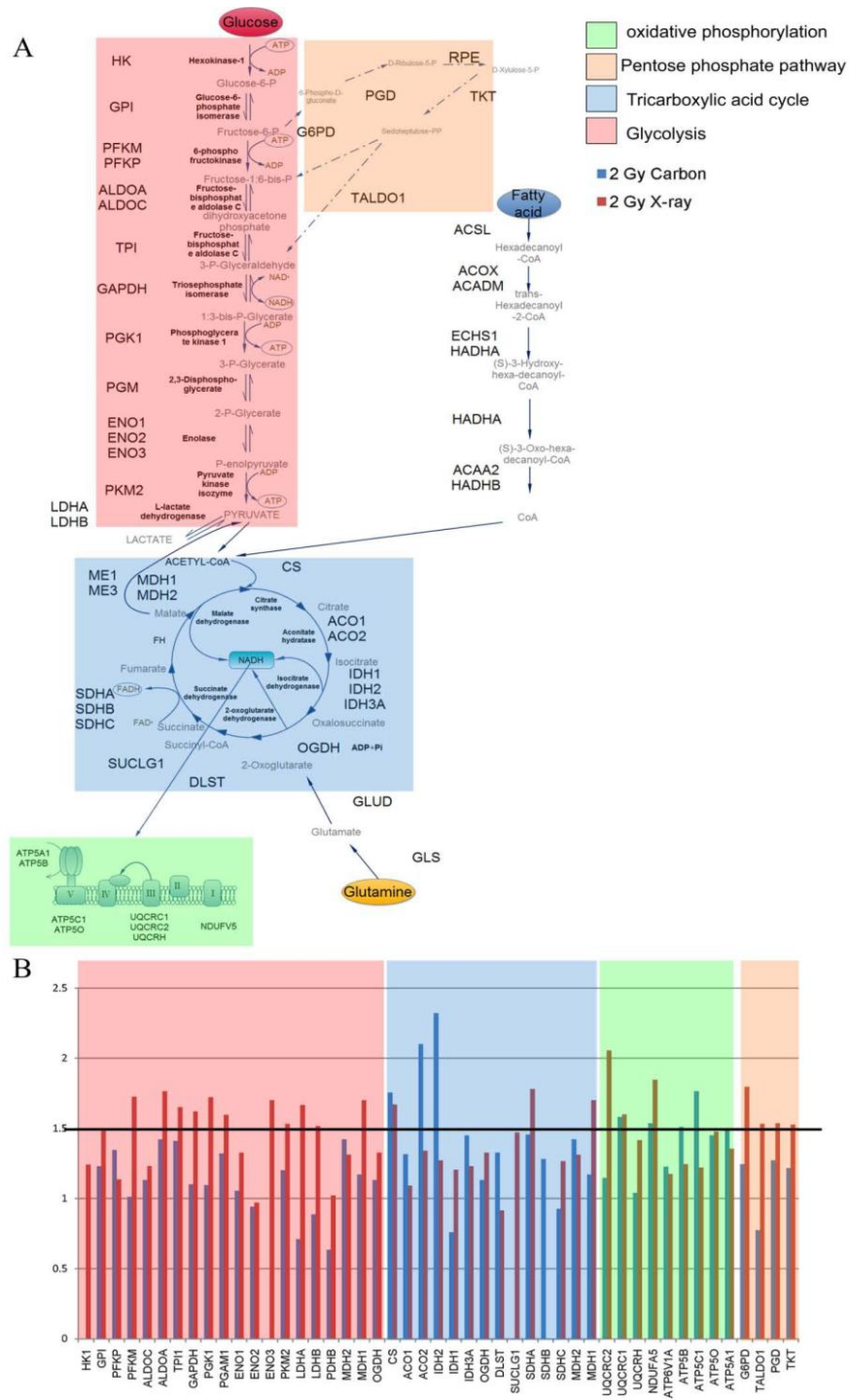

**Figure 3. Energy metabolic map of three differently irradiated samples.**

(A) Energy metabolic pathways. (B) The expression ratios of proteins associated with energy metabolism. The expression ratios with significant changes were separated by black line at 1.5 ratio.

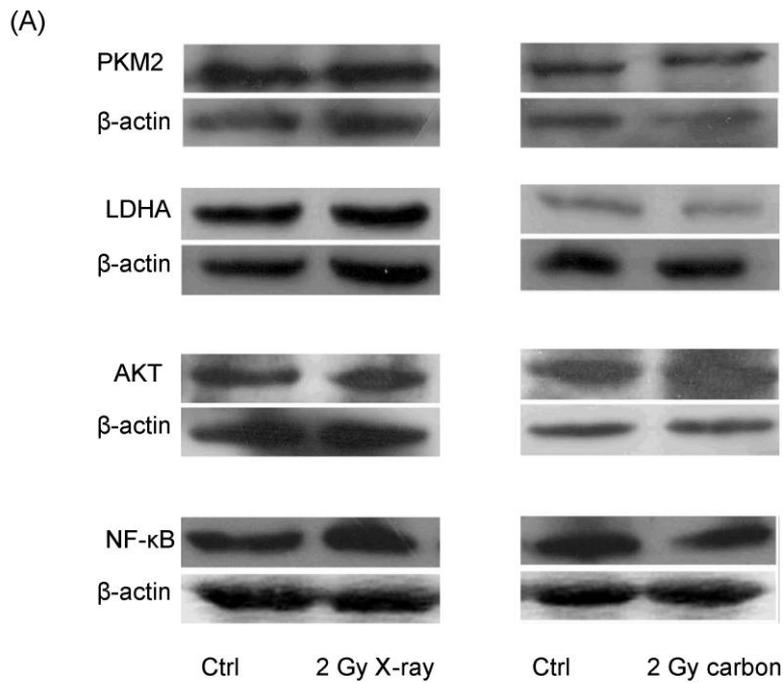

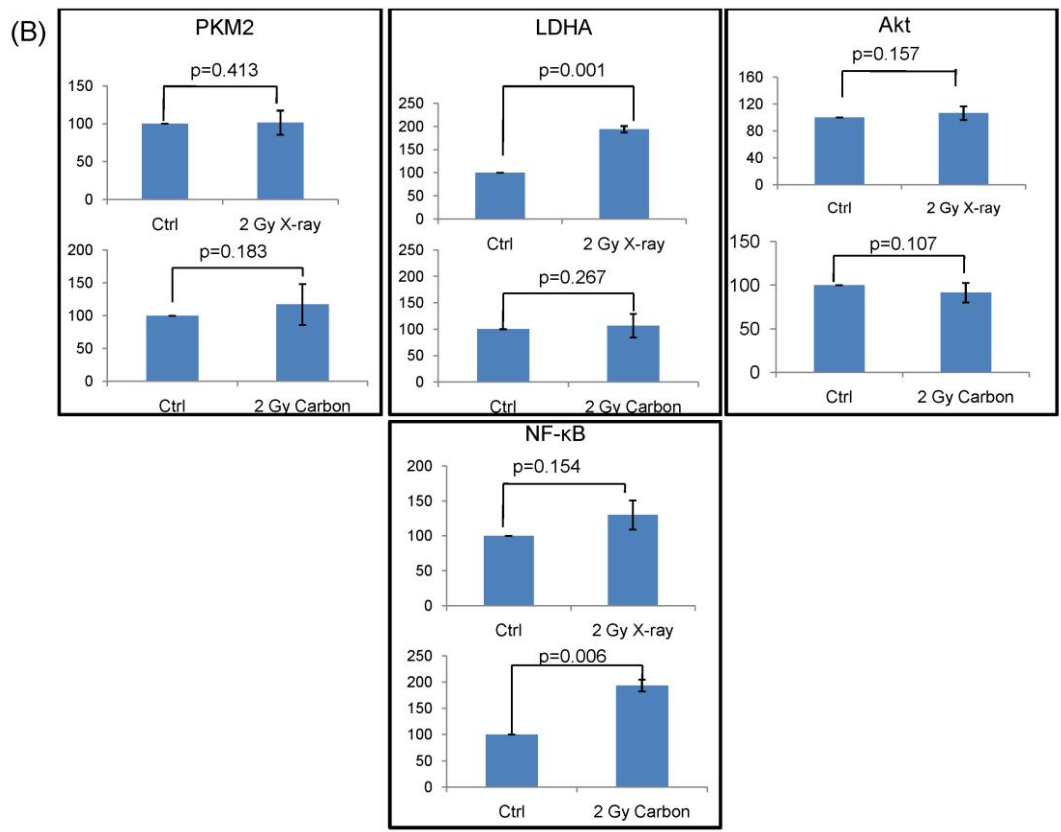

**Figure 4. Expression levels of PKM2, LDHA, AKT and NF-κB in HeLa cells 1 h after irradiation were measured with Western blot analysis and each experiment was applied by three replicates.**

(A) Expression levels of PKM2, LDHA, AKT and NF-κB in HeLa cells 1 h after irradiation were measured with Western blot analysis and each experiment was applied by three replicates. (B) A representative Western blot result for PKM2, LDHA, Akt and NF-κB. Expression of these proteins was presented as the relative grey intensity of the Western results from three independent repeats (The t-tests is one-tailed and p<0.05 was considered as significant).

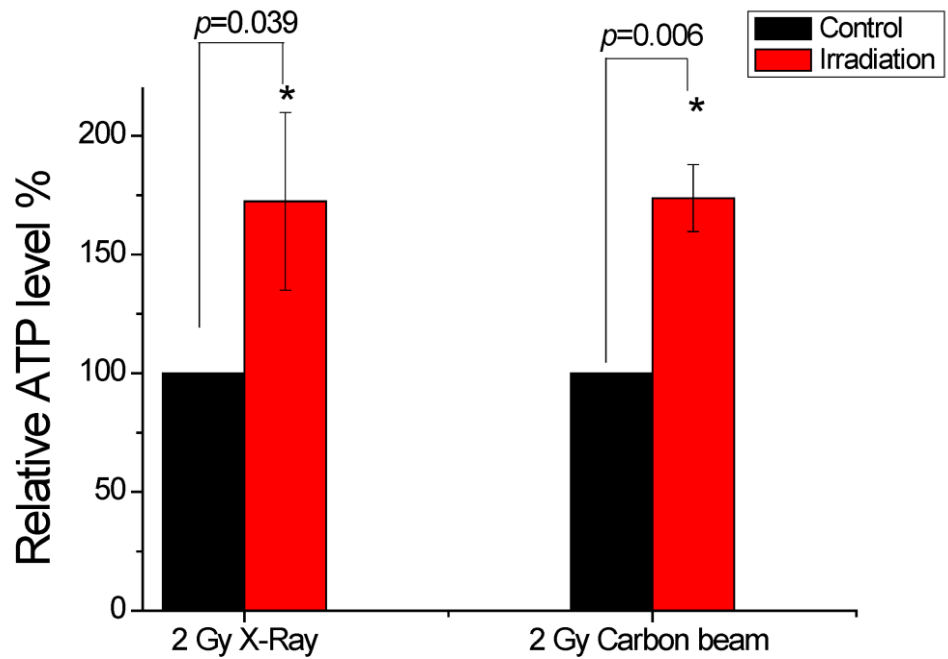

**Figure 5. Changes in ATP levels between cells exposed to different irradiation and sham control (\*, *p*<0.05, t-test was one-tailed)**.

One hour after irradiated, cells were harvested and submitted to ATP measurement. Data were normalized to sham control and independently repeated at least three times.

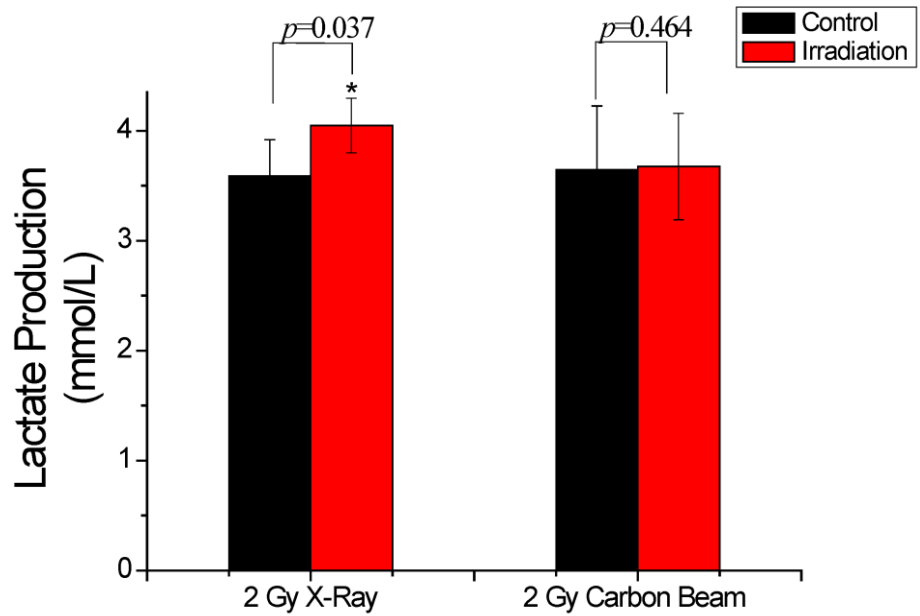

**Figure 6. Differences in lactic acid levels between sham control and irradiated cells (\*, *p*<0.05, t-test was one-tailed).**

One hour after irradiated, cells were harvested and Lactate was measured with a commercial kit by following manufacturer's instruction.

# Table Legends

**Table 1** Proteins involved in Energy metabolism process from DAVID

| Energy metabolism process | Gene Name | *p* value | Fold Enrichment |
|---|---|---|---|
| GG | LDHB, LDHA, ALDOC, PGAM1, PFKP, PFKM, OGDH, PDHB, GPI, TPI1, PKM2, ENO2, PGK1, GAPDH, MDH2, ENO1, MDH1 | 4.65E-11 | 8.43 |
| PPP | TPI1, TALDO1, PGD, HIBADH | 5.42E-3 | 10.35 |
| TCAC | SDHA, KYNU, ACO2, ACO1, IDH2, IDH1, IDH3A, PDHB, MDH2, MDH1 | 3.01E-6 | 9.11 |
| OP | UQCRC2, ATP6V1A, NDUFA5, UQCRC1, UQCRH, ATP5B, ATP5C1, ATP5O, ATP5A1 | 5.91E-2 | 2.14 |
| GS | CTPS, ALDH5A1, GLS, GGH, PHGDH, CAD, GMPS, PFAS | 8.24E-6 | 9.81 |
| FAM | PTGES3, ACAA2, PTGES2, ALDH5A1, ACOT2, LYPLA1, HADHA, HADHB, AKR1C3, TPI1, FASN, LTA4H, HSD17B4, ACSL3, RNPEP | 4.58E-2 | 1.76 |

**Table 2** Gene expression measured with Western blot assay and SILAC analysis

| Gene Name | Normalized Ratio | | Mean Ratio of SILAC | |
|---|---|---|---|---|
| | 2 Gy X-ray/Ctrl | 2 Gy carbon/Ctrl | 2 Gy X-ray/Ctrl | 2 Gy carbon/Ctrl |
| PKM2 | 1.01 ±0.18 | 1.16 ±0.21 | 1.53 ±0.06 | 1.20 ±0.36 |
| LDHA | 1.93 ±0.07 | 1.06 ±0.22 | 1.67 ±0.57 | 0.71 ±0.06 |
| Akt | 0.83 ±0.00 | 0.92 ±0.01 | -- | -- |
| NF-κB | 1.29 ±0.04 | 1.93 ±0.02 | -- | -- |